\def\D{\hbox{D\kern-.73em\raise.25ex\hbox{-}\raise-.25ex\hbox{ }}}
\def\d{\hbox{d\kern-.33em\raise.75ex\hbox{-}\raise-.75ex\hbox{}}}
\documentclass{llncs}
\usepackage{amssymb}
\usepackage[dvips]{graphicx}

\begin{document}

\pagestyle{empty}

\mainmatter

\title{A Method for Clustering Web Attacks \\ Using Edit Distance}


\author{Slobodan Petrovi\'{c}
\and Gonzalo \'{A}lvarez
}


\institute{Instituto de F\'{\i}sica Aplicada,
Serrano 144,\\
28006 Madrid, Spain\\
\email{\{slobodan, gonzalo\}@iec.csic.es}
}

\maketitle

\begin{abstract}
Cluster analysis often serves as the initial step in the process of data classification.
In this paper, the problem of clustering different length input data is considered. The edit
distance as the minimum number of
elementary edit operations needed to transform one vector into another is used. A heuristic
for clustering unequal length vectors, analogue to the well known $k$-means algorithm is
described and analyzed. This heuristic determines cluster centroids expanding shorter
vectors to the lengths of the longest ones in each cluster in a specific way. It is shown that
the time and
space complexities of the heuristic are linear in the number of input vectors. Experimental
results on real data
originating from a system for classification of Web attacks are given.
\end{abstract}

\section{Introduction}

Clustering can be informally defined as the process of grouping objects
that are similar in some way, where the number of groups may be unknown.
The proces of classification often begins with clustering of a data subset,
in order to determine the initial categories. The clusters obtained in
such a way are then used to categorize the rest of available data.
The methods of cluster analysis of equal length vectors have been widely treated in the
literature
(see, for example, \cite{Anderberg,FraRaf,Fukunaga,Hartigan}). For such clustering,
the classical distance measures can be used, such as the well
known Hamming distance. However, if unequal length vectors are to
be clustered, it is necessary to introduce new distance measures, since the classical
ones cannot be used in such cases \cite{Sankof}.

Two major groups of clustering methods exist: hierarchical (agglomerative), in which each
group of size greater than one is composed of smaller groups, and non hierarchical
(partitioning),
in which every object is assigned to exactly one group.

In this paper, a non hierarchical heuristic for clustering vectors of different lengths,
analogue to the $k$-means procedure of MacQueen \cite{MacQueen} is described and analyzed.
The unconstrained
edit distance
measure, as the minimum number of elementary edit operations (deletions and
substitutions) needed to transform one vector into another is used as a distance measure.
The essence
of the heuristic is the method of generating the new centroids. This is performed by
expanding the shorter vectors of each current cluster to the length of the longest one in
that cluster and manipulating the numbers of symbol occurences at each coordinate of these
expanded vectors. The time and memory complexities of the heuristic are analyzed, and
experimental
results on artificial as well as real data (the encodings of some Web attacks) are given.

The paper is organized as follows. Section 2 gives some preliminaries about clustering in
general and
the unconstrained edit distance. In Section 3, the new heuristic for
clustering vectors of different lengths analogue to the $k$-means algorithm is described,
together with
the possible variants, obtained by modifying the way of manipulating the numbers
of symbol occurences at the coordinates of expanded vectors.
In Section 4, the time and space complexities of the heuristic are analyzed. Finally,
in section 5,
the experimental results on random as well as real samples are given.

\section{Preliminaries}

In this paper, we consider the following problem:

Let ${\cal P}$ be a set of vectors, whose cardinality is $m$, and whose elements are
$\mathbf{p}_{1},\ldots ,\mathbf{p}_{m}$, of dimensions $n_{1},\ldots , n_{m}$,
respectively. The task of cluster
analysis of such vectors is:
partition the set ${\cal P}$ into $k$ nonempty subsets, $P_{1},\ldots ,P_{k}$, such that the
following holds:

\begin{equation}
P_{1}\cup P_{2}\cup\ldots\cup P_{k}={\cal P}
\end{equation}

\begin{equation}
P_{i}\cap P_{j}=\varnothing,\hphantom{xxx}i,j=1,2,\ldots ,k,\hphantom{xxx}i\not = j,
\end{equation}

\noindent
optimizing some of the partition criteria. The number $k$ may be given in advance,
but it need not be.
The partition criterion that is usually used is the {\em sum-of-squares criterion}.

Let $\mathbf{p}_{i}^{r}$ and $\mathbf{p}_{i}^{s}$ be the $r$-th and $s$-th element
of the subset $P_{i}$ of the
set ${\cal P}$, $i\in\{1,\ldots ,k\}$, $r,s\in\{1,\ldots ,\mid P_{i}\mid\}$, $r\not = s$. Let
$d(\mathbf{p}_{i}^{r},\mathbf{p}_{i}^{s})$ be the distance between $\mathbf{p}_{i}^{r}$
and $\mathbf{p}_{i}^{s}$ defined in some
way. Then the sum-of-squares criterion is given by the following expression:

\begin{equation}\label{eq:partcrit1}
c(P_{i})=\sum_{r,s=1;r\not = s}^{\mid P_{i}\mid}(d(\mathbf{p}_{i}^{r},
\mathbf{p}_{i}^{s}))^{2}.
\end{equation}

It is well known (see for example \cite{GareyJohnson}) that the problem of minimization of
the sum-of-squares criterion is NP hard. However, there are many algorithms that find a local
minimum of this criterion. For example, the well known $k$-means algorithm \cite{MacQueen} is
of this kind.

Let $\mathbf X$ and $\mathbf Y$ be vectors of lengths $N$ and $M$, respectively,
whose coordinates take the
values from the discrete alphabet ${\cal A}$. Edit distance between the vectors $\mathbf X$
and $\mathbf Y$ is
defined as the minimum number of elementary edit operations (substitutions, deletions,
and/or insertions) needed to transform $\mathbf X$ into $\mathbf Y$. Certain constraints
can be incorporated into
this definition, which can concern the total number of elementary edit operations, the maximum
deletion and/or insertion run lengths, the total number of deletion and/or insertion runs, etc.
The combinations of these constraints are also possible \cite{PetGol,Sankof}.
In this paper, the unconstrained edit distance, where the elementary edit operations are
substitutions and deletions, is used as it models well the possible transformations of the
input vectors.

Nonnegative real-valued elementary edit distances are associated with the corresponding
elementary edit operations:

\begin{list}{}{\partopsep -2mm\parsep 0mm\itemsep 2mm\leftmargin 5mm}
\item[1.] $d(x,\phi)$ is the elementary distance associated
with the deletion of $x \in {\cal A}$ from the vector $\mathbf X$,
where the 'empty' symbol $\phi$ is introduced to represent deletion;
\item[2.] $d(x,y)$ is the elementary distance associated with
the substitution of $x$ by $y$, $x,y \in {\cal A}$. Usually, $d(x,x)=0, \forall x$.
If $d(x,y)\not = 0$,
then the corresponding substitution is called {\em the effective substitution}.
\end{list}

In order to define the explicit expression for the constrained
edit distance, an edit transformation can be represented sequentially\@.
Namely, we define a 2-dimensional
edit sequence ${\cal E}=([\alpha],[\beta])$
over the alphabet $\{0,1,\phi\}$
by the following
encoding scheme.

First, let for an arbitrary vector $\mathbf G$
over ${\cal A}$, $\mathbf{\gamma}$ denote any vector over $\{0,1,\phi\}$
such that by
removing all the 'empty'
symbols from $\mathbf{\gamma}$ one obtains $\mathbf G$. Then, an edit-sequence
$([\alpha],[\beta])$
is defined applying the following rules:

\begin{list}{}{\partopsep -2mm\parsep 0mm\itemsep 2mm\leftmargin 5mm}
\item[1.] The lengths of $[\alpha]$ and $[\beta]$
are equal to $N$.
\item[2.] If $\alpha (i)$ and $\beta (i)$ are non-empty symbols, then the
substitution of the symbol $\alpha (i)$ by the symbol $\beta (i)$ takes place,
for any $1\leq i \leq N$.
\item[3.] If $\alpha (i)$ is not the 'empty' symbol, and $\beta (i)$ is the
'empty' symbol,
then the deletion of $\alpha (i)$ takes place, for any $1\leq i \leq N$.
\item[4.] For any $1 \leq i \leq N$ no other cases apart from
2.  and 3. are allowed.
\end{list}

There is an one-to-one correspondence
between
the set of all the permitted edit sequences
$([\alpha],[\beta])$
defined as
above,
denoted by $\Gamma ({\mathbf X},{\mathbf Y})$,
and the set of all the permitted
edit transformations of $\mathbf X$ into $\mathbf Y$\@.

The edit distance can be expressed in terms of edit sequences
by:

\begin{equation}
D({\mathbf X},{\mathbf Y}) =
\min\left\{\sum_{i=1}^N
d(\alpha (i),\beta (i))\mid
([\alpha],[\beta]) \in \Gamma({\mathbf X},{\mathbf Y})
\vphantom{\sum_{i=1}^{\cal L}}\right\}
\end{equation}

{\bf Example:} Let
${\mathbf X}=(1,0,1,1,0,1,1,0,0,0,1)$ and
${\mathbf Y}=(1,0,1,1,1,1,0)$. An edit sequence that corresponds to a permitted
edit transformation is given by

\begin{center}
\hspace*{2mm}${\cal E}=\left[\vphantom{\begin{array}{l}
\phi \\
\phi
\end{array}}\right.$
\begin{tabular}{@{\extracolsep{1.4mm}}l@{}l@{}l@{}l@{}l@{}l@{}l@{}l@{}l@{}l@{}l@{}l@{}}
1&0&1&1&0&1&1&0&0&0&1& \\
1&0&$\phi$&$\phi$&1&1&1&$\phi$&1&0&$\phi$&
\end{tabular}
\hspace*{-2mm}$\left.\vphantom{\begin{array}{l}
\phi \\
\phi
\end{array}}\right]$.
\end{center}

Assuming that the elementary distances associated with deletions and
effective substitutions are all equal to one, by using (4), one can
determine that the edit distance corresponding to this edit sequence is $6$\@.


The unconstrained edit distance can be calculated recursively, by filling the matrix of
partial edit distances \cite{Oommen}. Let $\mathbf X$ and $\mathbf Y$ be vectors of
lengths $N$ and
$M$, respectively, over the finite alphabet ${\cal A}$. Let $e$ be the number of
deletions and let $s$ be the number of substitutions in an edit transformation of the
prefix $\mathbf{X}_{e+s}$ of $\mathbf X$ to the prefix $\mathbf{Y}_{s}$ of $\mathbf Y$.
Let $d(x,y)$ be the
elementary edit distance associated with the substitution of the symbol $x$ by
the symbol $y$ and let $d(x,\phi)$ be the elementary edit distance associated with the
deletion of the symbol $x$ from $\mathbf X$. Then the partial edit distance
$W[e,s]$ between $\mathbf{X}_{e+s}$ and $\mathbf{Y}_{s}$ can be computed following
the same lines as in \cite{Oommen}:

\begin{displaymath}
W[e,s]=\min\left\{\begin{array}{l}
W[e,s-1]+d(X(e+s),Y(s)), \\
W[e-1,s]+d(X(e+s),\phi)
\end{array}\right.
\end{displaymath}

\begin{equation}
e=1,\ldots ,N\hphantom{x}s=1,\ldots ,\min\{N-e,M\}.
\end{equation}

In the sequel, it will be assumed that:

\begin{list}{}{}
\item[1.] $d(x,\phi)=d_{e},\forall x\in {\cal A}$.
\item[2.] $d(x,x)=0,\forall x\in {\cal A}$.
\end{list}

The following algorithm implements the relation (5) in order to determine the edit
distance between the given vectors of different lengths.

{\bf Algorithm 1}

{\bf INPUT:}

The vectors $\mathbf X$ and $\mathbf Y$ of lengths $N$ and $M$, respectively.

Elementary edit distance $d_{e}$ associated with the deletion of a symbol
\mbox{from $\mathbf X$}.

Elementary edit distance $d(x,y)$ associated with the substitution of the
symbol $x$ by the symbol $y$, $\forall x,y$.

{\bf OUTPUT:}

The matrix $\mathbf W$ of partial edit distances associated with the transformations
of the prefixes of the vector $\mathbf X$ to the corresponding prefixes of the vector
$\mathbf Y$.

{\tt // Initialization:}

{\tt W[0][0] = 0 ;}

{\tt // The column 0 of the matrix W:}

{\tt for e = 1 to N}

\hspace*{0.5cm}$\texttt{W[e][0] = W[e-1][0] + d}_{\texttt{e}} \texttt{;}$

{\tt // The row 0 of the matrix W:}

{\tt for s = 1 to M}

\hspace*{0.5cm}{\tt W[0][s] = W[0][s-1] + d(X[s],Y[s]) ;}

{\tt // Main loop:}

{\tt for e = 1 to N}

\hspace*{0.5cm}{\tt for s = 1 to min(N-e,M)}

\hspace*{0.7cm}{\tt W[e][s] = min}$\texttt{(W[e-1][s] + d}_{\texttt{e}}\texttt{, W[e][s-1] +
d(X[e+s],Y[s])) ;}$

{\tt // Calculate the edit-distance:}

{\tt d(X,Y) = W[N-M][M] .}


It is possible to reconstruct one of the possible optimal edit sequences, either by
backtracking
through
the matrix of partial edit distances $\mathbf W$ or by maintaining pointers to the cells of
$\mathbf W$ during
the execution of the Algorithm 1. The reconstructed sequence is not unique in general. The
following algorithm reconstructs one of the possible optimal edit sequences that transforms the
vector $\mathbf X$ of length $N$ into the vector $\mathbf Y$ of length $M$, by backtracking
through
the \mbox{matrix $\mathbf W$}.

{\bf Algorithm 2}

{\bf INPUT:}

The vectors $\mathbf X$ and $\mathbf Y$ of lengths $N$ and $M$, respectively.

Elementary edit distance $d_{e}$ associated with the deletion of a symbol
\mbox{from $\mathbf X$}.

The matrix $\mathbf W$ of partial edit distances between the prefixes of the vector
$\mathbf X$ and the corresponding prefixes of the vector $\mathbf Y$.

{\bf OUTPUT:}

One of the optimal edit sequences $([\alpha],[\beta])$ that transforms $\mathbf X$
into $\mathbf Y$.

{\tt // Initialization:}

{\tt e = N-M ;}
{\tt s = M ;}

{\tt L = 0 ; // The length of the edit sequence}

{\tt // Backtracking through the matrix W:}

{\tt while ((e>0) or (s>0))}$\{$

\hspace*{0.5cm}{\tt if} $\texttt{(W[e][s] == W[e-1][s] + d}_{\texttt{e}}\texttt{)}\{$

\hspace*{1cm}{\tt L++ ;}

\hspace*{1cm}{\tt alpha[L] = X[e+s] ;}

\hspace*{1cm}{\tt beta[L] = }$\varnothing${\tt ;}

\hspace*{1cm}{\tt e-- ;}

\hspace*{0.5cm}$\}$

\hspace*{0.5cm}{\tt else}$\{$

\hspace*{1cm}{\tt L++ ;}

\hspace*{1cm}{\tt alpha[L] = X[e+s] ;}

\hspace*{1cm}{\tt beta[L] = Y[s] ;}

\hspace*{1cm}{\tt s-- ;}

\hspace*{0.5cm}$\}$

$\}$


\section{The new heuristic}

In this section we describe the new heuristic for clustering unequal length vectors, analogue
to the $k$-means algorithm.

Let ${\cal P}$ be a set of vectors, whose cardinality is $m$, and whose elements are
$\mathbf{p}_{1},\ldots ,\mathbf{p}_{m}$, of dimensions $n_{1},\ldots , n_{m}$,
respectively. Let the coordinates of the vectors
take values from the discrete finite alphabet ${\cal A}$. Let $k$ be the number of clusters
given in advance. The heuristic starts from an arbitrary initial partition of ${\cal P}$
into $k$ clusters, $P_{1},\ldots ,P_{k}$. Since the vectors in the clusters are of
different lengths, the heuristic expands the shorter vectors in every
cluster to the length of the longest one in the same cluster. This is carried out by means
of the optimal edit sequences that transform the longest vector in the cluster to each of
the remaining vectors in it. Then the coordinates of the cluster's centroid are calculated
counting
the symbols at the corresponding coordinates of the expanded vectors and selecting the most
frequent symbol at each coordinate to be the symbol at the corresponding coordinate of the
new centroid. Note that the obtained centroid is expanded in general, since it can contain
empty symbols. The final new centroid of the cluster is obtained by removing the empty
symbols from the expanded centroid.

The new clusters are created by finding the minimum value of the unconstrained edit
distance
between each member of ${\cal P}$ and the new centroids. The process continues until the new
clusters are equal to the previous ones.

The following is the formal description of the heuristic.

{\bf Algorithm 3:}

{\bf INPUT:}

The set ${\cal P}$ of vectors $\mathbf{p}_{1},\ldots ,\mathbf{p}_{m}$ of lengths
$n_{1},\ldots , n_{m}$, respectively.

The number of clusters $k$.

{\bf OUTPUT:}

A partition of the set ${\cal P}$ into $k$ clusters.

{\tt // Initialization:}

{\tt terminate = false ;}

Select the initial centroids $C_{1},\ldots ,C_{k}$. These could be, for example,
any $k$ vectors from the input set ${\cal P}$, chosen at random.

Calculate the unconstrained edit distance between every vector from ${\cal P}$ and
every centroid from the set ${\cal C}=\{C_{1},\ldots ,C_{k}\}$, using the Algorithm 1.
Assign every vector from ${\cal P}$ to the nearest centroid. In such a way,
the initial clustering $P_{1},\ldots ,P_{k}$ is obtained.

{\tt Main loop:}

{\tt while(not terminate)}$\{$

\hspace*{0.5cm}{\tt // Calculate the new centroids:}

\vspace*{0.2cm}
\hspace*{0.5cm}
\hfill\begin{minipage}{11cm}
Let $\mathbf{p}_{i}^{s}$ be the longest vector in the cluster $P_{i}$, $i=1,\ldots k$.
Let $\mathbf{p}_{i}^{r}$, $r\in\{1,\ldots ,\mid P_{i}\mid\}\setminus \{s\}$ be other
elements of the cluster $P_{i}$.

For each $r\in\{1,\ldots ,\mid P_{i}\mid\}\setminus \{s\}$, find an optimal
edit sequence $([\alpha_{i}^{r}],[\beta_{i}^{r}])$ that transforms $\mathbf{p}_{i}^{s}$ to
$\mathbf{p}_{i}^{r}$, using the Algorithm 1 and the Algorithm 2.
The $[\beta_{i}^{r}]$ is the expanded vector $\mathbf{p}_{i}^{r}$.

Find the symbol that prevails at every coordinate of the expanded vectors
$[\beta_{i}^{r}]$,
$r\in\{1,\ldots ,\mid P_{i}\mid\}\setminus \{s\}$. Make this symbol the
new value of the corresponding coordinate of the new expanded centroid $C_{i}''$
of the cluster $P_{i}$.

Remove the empty symbols from the expanded centroid $C_{i}''$. In such a way, the
new centroids $C_{i}'$, $i=1,\ldots k$ are obtained.
\end{minipage}

\vspace{0.2cm}
\hspace*{0.5cm}{\tt // Reassign the input vectors to the new centroids:}

\vspace{0.2cm}
\hfill\begin{minipage}{11cm}
Assign every input vector from ${\cal P}$ to the nearest centroid from the set
${\cal C}'=\{C_{1}',\ldots ,C_{k}'\}$, by calculating the edit distance between the
vectors and the new centroids, using the Algorithm 1. Thus the new clustering
$P_{1}',\ldots ,P_{k}'$ is obtained.
\end{minipage}

\hspace*{0.5cm}{\tt // Check if the new clustering is equal to the previous one:}

\hspace*{0.5cm}{\tt if $\texttt{((P}_{\texttt{1}}\texttt{,\ldots P}_{\texttt{k}}
\texttt{) == (P}_{\texttt{1}}\texttt{',\ldots ,P}_{\texttt{k}}\texttt{')}\texttt{)}$}

\hspace*{1cm}{\tt terminate = true ;}

\hspace*{0.5cm}{\tt else}$\{$

\hspace*{1cm}{\tt $\texttt{(P}_{\texttt{1}}\texttt{,\ldots ,P}_{\texttt{k}}
\texttt{) = (P}_{\texttt{1}}\texttt{',\ldots ,P}_{\texttt{k}}\texttt{')}$ ;}

\hspace*{1cm}{\tt ${\cal C} = {\cal C}'$ ;}

\hspace*{0.5cm}$\}$

$\}$


In the Algorithm 3, the new centroids of the clusters are obtained by counting the symbols
at the
coordinates of the expanded vectors and selecting the symbol that prevails. There might,
however,
be the cases in which all the possible symbols occur equal number of times at one of the
coordinates. Possible solutions of the problem in such cases are:

\begin{list}{}{}
\item[1.] Choose the symbol at random, among those present at the coordinate;
\item[2.] Choose the symbol whose position in the alphabet is the closest to the first symbol;
\item[3.] Choose the symbol whose position in the alphabet is the closest to the last symbol;
\item[4.] Choose the empty symbol, if present at the coordinate.
\end{list}

Obviously, the most objective results are obtained by selecting the symbol at those coordinates at
random. But by using other variants, one can fine-tune the heuristic favorizing various
types of new centroids. For example, by choosing the empty symbol in such cases, the shorter
centroids are favorized, since the empty symbols are dropped from the expanded centroids.

It is easy to see that the coordinates of the input vectors can take values from different
alphabets. This can be of particular importance in the applications. In that case, the elementary
edit distances can be defined (although it is not obligatory) in such a way that the substitutions
of the symbols from the same alphabet are favorized.

\section{The complexity of the heuristic}

It can be shown (see for example \cite{Anderberg}) that both time and space complexities of
the $k$-means algorithm are ${\cal O}(m)$, where $m$ is the number of input vectors.

The new heuristic described in this paper basically consists of the same steps as the $k$-means
algorithm. Let $m$ be the number of vectors of dimensions $n_{1},\ldots ,n_{m}$ in the input
data set of the heuristic.
Let $k$ be the number of clusters and let $R$ be the number of iterations (i.e.
the number of times new centroids are calculated) of the heuristic. In the $i$-th
iteration, let $m_{1}^{i},\ldots ,m_{k}^{i}$ be the cardinalities of the clusters,
$i=1,\ldots ,R$. In every iteration, prior to determining of the symbols that prevail at the
coordinates of the expanded vectors in the cluster, these expanded vectors must be obtained by
means of the Algorithms 1 and 2. The complexity of these algorithms is quadratic in the length
of the input sequences \cite{Oommen}. Let $n_{\mathrm{max}}$ be the length of the longest vector
in the input data set. Then the number of operations needed to transform all the vectors in all
the clusters to the longest vector of their corresponding cluster is
$\sim mn_{\mathrm{max}}^{2}$, since the elementary edit operations are only deletions and
substitutions,
thus making the lengths of all the edit secuences $\leq n_{\mathrm{max}}$.

To determine the new centroids starting from the expanded vectors,
$\sim (m_{1}^{i}+\cdots +m_{k}^{i})n_{\mathrm{max}}$ operations are needed in each iteration,
$i=1,\ldots R$. Since by definition the clusters are mutually exclusive,
the number of operations needed to determine the new centroids of all the clusters is
$\sim mn_{\mathrm{max}}$.

The
number of operations needed to assign all the input vectors to the nearest centroid is
$\sim kmn_{\mathrm{max}}^{2}$. Thus the total number of operations of the heuristic is
$\sim Rm(n_{\mathrm{max}}^{2}(1+k)+n_{\mathrm{max}})$, which means that the time complexity of
the heuristic is $O(m)$.

In \cite{Anderberg} the convergence properties
of the $k$-means algorithm have been studied experimentally and the conclusion was given that
the expected number of
iterations was very small. Having in mind the essential similarity between this heuristic and
the $k$-means algorithm,
similar results to those from \cite{Anderberg} concerning the convergence can be expected.

The storage needed by the heuristic is $\sim n_{max}m$ memory cells to store the input
vectors, and $\sim n_{max}^{2}$ memory cells needed to store the matrix of partial edit
distances.
Thus the space complexity of the heuristic is $O(m)$.

\section{Experimental results}

The heuristic described in this paper was tested in two ways:

\begin{list}{}{}

\item[1.] A set of artificially generated discrete vectors
was clustered by means of the heuristic. 1000 random samples were
generated, consisting of 2000 vectors of
length at most 20, naturally grouped into 2 clusters. In 40\% of these examples, there was
no overlapping between clusters at all. In 30\% of the examples there was 10\% of
overlapping between clusters. Finally, in 30\% of the examples there was 20\% of
overlapping between clusters. The number of
incorrectly clustered vectors was assigned to categories. The results obtained with the
heuristic
are presented in the Fig. 1. As it can be seen, the cases without
overlapping were resolved correctly, whereas in the cases with overlapping the
results obtained with the heuristic depended directly on the
overlapping degree. This behaviour of the heuristic was similar to the behaviour of the
$k$-means heuristic which is known to be sensitive to
overlapping \cite{Fasulo}.

\item[2.] A data set originating from the system whose intention was to classify
the attacks on a Web server into a number of categories was clustered by means of the
heuristic. In this system, there is a need to cluster a
substantial amount of vectors of different lengths that describe the security
alerts. The encoding scheme of these alerts is given in \cite{AlvPet}. The input vectors have
discrete coordinates that generally take values from different alphabets. The encoding
given in \cite{AlvPet} recognizes 9 properties,
which means that the maximal length
of a vector from the input data set can be 9. But the number of input vectors can be very large.
The correctness of the clustering was checked against the values of the "severity"
indicator of
the "SNORT" intrusion detection system \cite{Snort}. 1000
samples originating from the system were tested, consisting of 5000 vectors of length at most 9,
where the number of clusters (according to the "severity" value defined in the "SNORT" system)
varied between 2 and 4. The number of incorrectly clustered vectors was assigned to
categories.
The results obtained with the heuristic are presented in the Fig. 2. As
it can be seen, the input vectors were clustered correctly in $\approx 75\%$ of the given
cases.

\end{list}



\begin{center}
\includegraphics[scale=.8]{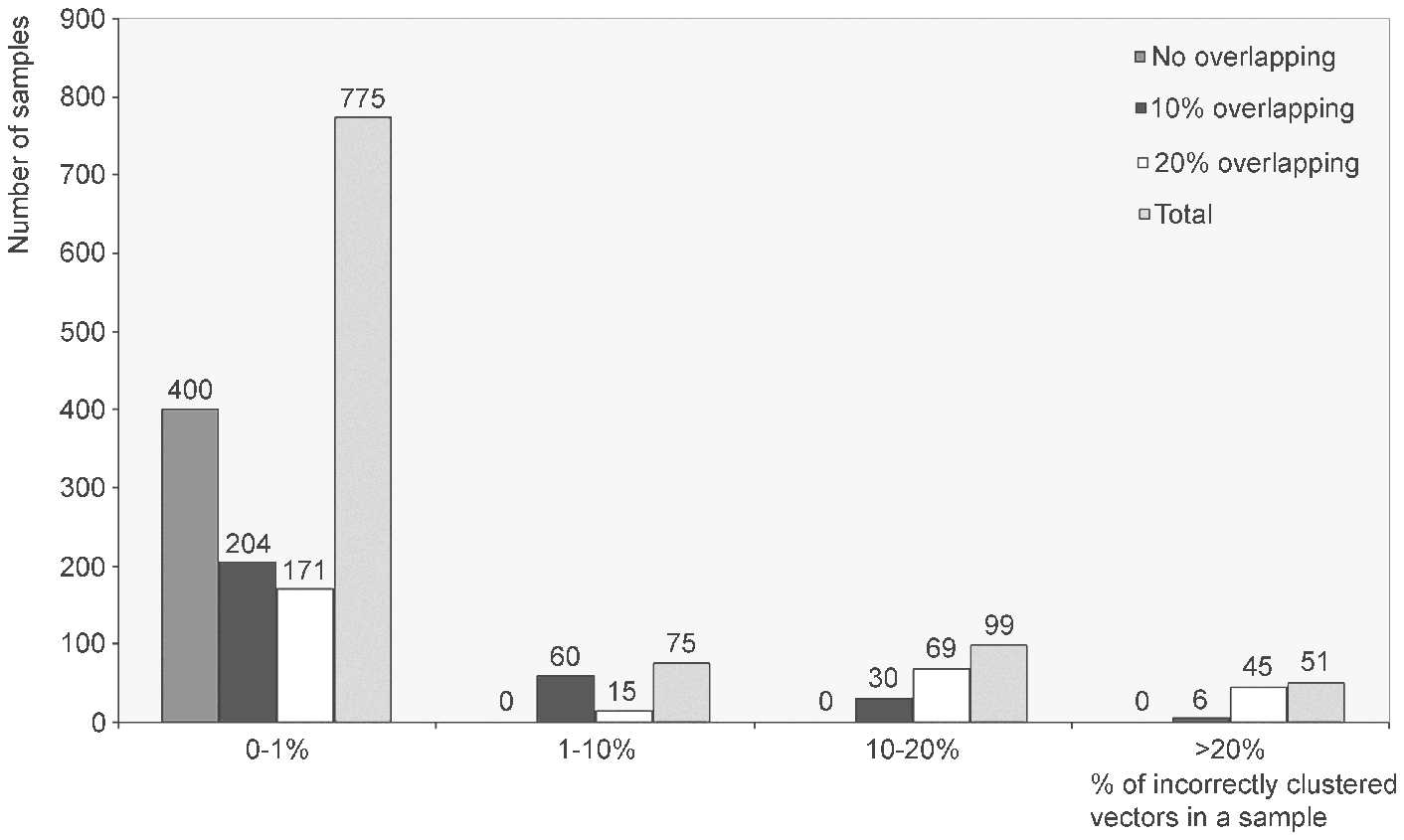}

{\em Fig. 1 - Results of clustering artificially generated
vectors}
\end{center}

\begin{center}
\includegraphics[scale=.8]{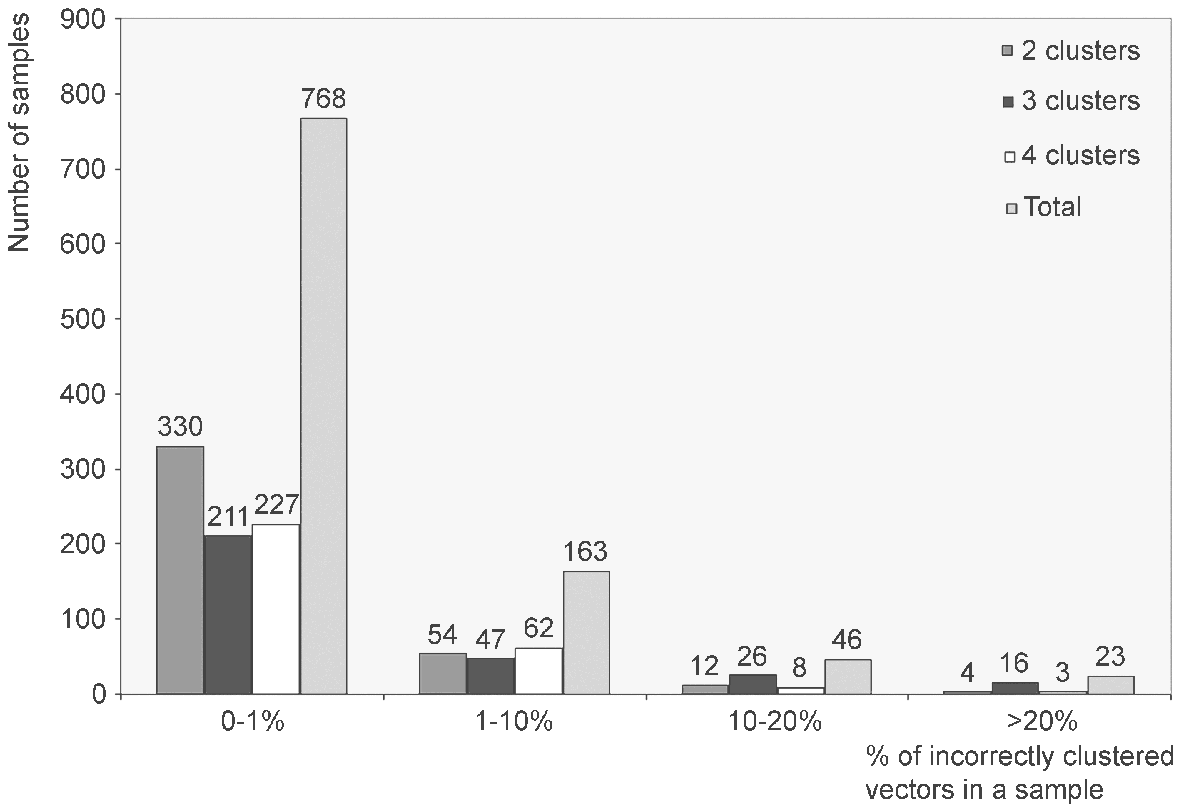}

{\em Fig. 2 - Results of clustering vectors from a real system}
\end{center}

\section{Conclusion}

In this paper, a new heuristic for clustering unequal length vectors with discrete coordinates is
described and analyzed. The heuristic is analogue to the $k$-means algorithm for
clustering equal
length real vectors. It expands the member vectors of the clusters to the lengths of the
longest vectors in those clusters, in order to determine the new centroids in each iteration. It
was shown that both time and space complexities of the heuristic are linear in the number of
input vectors. The experimental results show that the
behaviour of
the heuristic is similar to that of the $k$-means algorithm considering overlapping of
the clusters,
and that the level of correctness of clustering with the heuristic is promissing.

\end{document}